\documentclass{article}
\usepackage{spconf,amsmath,graphicx,hyperref}
\usepackage{comment}
\usepackage{mathtools}
\usepackage{subfig}
\usepackage{tabularx,booktabs,multirow}
\usepackage{float}
\usepackage{placeins}
\usepackage{dblfloatfix}
\usepackage{times}
\usepackage{latexsym}
\usepackage{cite}
\usepackage{xcolor}
\usepackage{amsfonts}
\usepackage{amssymb}  
\usepackage{algorithm}
\usepackage{algpseudocode}
\usepackage{makecell}

\title{An Effective Energy Mask-based Adversarial Evasion Attacks against
Misclassification in Speaker Recognition Systems}

\name{Chanwoo Park, Chanwoo Kim\textsuperscript{†}\thanks{{†} Corresponding author}}
\address{\textit{Department of Artificial Intelligence, Korea University, }\textit{Seoul, Republic of Korea}}

\begin{document}

\maketitle

\begin{abstract}
  Evasion attacks pose significant threats to AI systems, exploiting vulnerabilities in machine learning models to bypass detection mechanisms.
  The widespread use of voice data, including deepfakes, in promising future industries is currently hindered by insufficient legal frameworks.
  Adversarial attack methods have emerged as the most effective countermeasure against the indiscriminate use of such data.
  This research introduces masked energy perturbation (MEP), a novel approach using power spectrum for energy masking of original voice data.
  MEP applies masking to small energy regions in the frequency domain before generating adversarial perturbations, targeting areas less noticeable to the human auditory model.
  The study primarily employs advanced speaker recognition models, including ECAPA-TDNN and ResNet34, which have shown remarkable performance in speaker verification tasks.
  The proposed MEP method demonstrated strong performance in both audio quality and evasion effectiveness.
  The energy masking approach effectively minimizes the perceptual evaluation of speech quality (PESQ) degradation, indicating that minimal perceptual distortion occurs to the human listener despite the adversarial perturbations.
  Specifically, in the PESQ evaluation, the relative performance of the MEP method was 26.68\% when compared to the fast gradient sign method (FGSM) and iterative FGSM.
\end{abstract}

\section{Introduction}

Voice spoofing attacks, speaker verification vulnerabilities \cite{qadir2022voice, chouchane2021privacy, wu2016study}, and malicious voice data manipulation have emerged as critical security threats in the era of artificial intelligence and machine learning systems.
The rapid advancement of deep learning technologies, particularly in voice synthesis and deepfakes, has created unprecedented challenges in maintaining the integrity of voice-based security systems and biometric authentication mechanisms.

Voice-based security threats have reached unprecedented complexity, with targeted noise signals achieving 95\% attack success rates against speaker recognition systems \cite{baroughi2014additive}.
Critical sectors like healthcare documentation and banking biometrics face escalating risks from voice cloning-enabled unauthorized access \cite{czyzewski2024enhancing, arefin2024strengthening}.
Three primary adversarial attack vectors threaten ML systems: data poisoning, inference extraction, and evasion techniques exploiting model vulnerabilities.

AI-enhanced biometric systems introduce new attack surfaces where synthetic voices successfully mimic biometric characteristics.
Although deep neural networks outperform humans in cloned voice detection \cite{olateju2024combating}, high biometric similarity remains exploitable.
Telecommunications infrastructure faces compounded risks including service denial and data integrity breaches.
The masked energy perturbation (MEP) method pioneers psychoacoustically-informed attacks by selectively distorting high-energy speech components while preserving perceptual quality.

Benchmark tests on LibriSpeech \cite{panayotov2015librispeech} demonstrate MEP's superiority over elementwise perturbation, maintaining perceptual evaluation of speech quality (PESQ) scores \cite{rix2001perceptual} above 3.5 and superior signal-to-noise ratio (SNR).
Experimental results show 20\% higher attack success rates than conventional methods, with robust performance across noise environments and speaker variations.

Multilayered defense strategies combining real-time anomaly detection and adaptive response protocols emerge as critical countermeasures \cite{olateju2024combating}.
Encryption protocols for biometric data storage and multi-factor authentication integration are recommended for enhanced security \cite{dilmaghani2019privacy}.
The research community prioritizes developing adversarial-resistant models and standardized security benchmarks.

\section{Methodology}
\label{sec:method}

\subsection{Energy Mask}

For the adversarial speaker attack, we extracted 512 frequency bins using a Hann window with a 25 ms window size and a 12.5 ms frameshift, adhering to the overlap-add (OLA) constraint \cite{verhelst2000}.
All recordings were resampled to 16 kHz.
This allows for perfect reconstruction of the original signal without modifying the signal itself.
When applying masking techniques to small energy regions while introducing distortion only to high energy components, we first consider the energy distribution across time-frequency bins, taking into account the human auditory model.
The small energy masking (SEM) algorithm \cite{kim2020small} masks time-frequency bins whose energy falls below a certain threshold.
This threshold is randomly generated using a uniform distribution relative to the peak energy of each utterance.
For the unmasked components, the values are scaled to maintain the total sum of feature values through this masking procedure.
The energy $x[m,k]$ in each time-frequency bin is calculated using the following equation, where $m$ is the frame index, and $K$ is the fast fourier transform (FFT) size used within each short-time fourier transform (STFT) window, and $S[m, e^{j \omega_k}]$ denotes the complex STFT coefficient:

\begin{equation}
  x[m,k] = |S[m, e^{j \omega_k} ] |^2 ,
\label{eq:1}
\end{equation}

where $\omega_k$ represents the discrete-time frequency, calculated as $\omega_k = \frac{2 \pi k}{K}$, where $k$ ranges from 0 to $K-1$.
The $x_{mask}$ value is determined by taking the highest value after excluding the top 5\% of energy $x_{peak}$ values when sorted in descending order for each utterance.
Since peaks are calculated within a single utterance, the actual values may vary for each utterance.
For a specific energy value $x[m,k]$, we can define the term $\eta$ and function $f$, which represents the ratio between $x[m,k]$ and the peak in decibels (dB).

\begin{equation}
  \eta = f\left(x[m,k]\right) := 10 \log_{10}\left(\frac{x[m,k]}{x_{peak}}\right)\,.
  \label{eq:2}
\end{equation}

More specifically, the masking-related procedure is explained within the dotted rectangle.
The energy $x[m,k]$ for each time-frequency bin is calculated using \eqref{eq:1}.

For each utterance, we calculate the maximum value of energy peaks $x_{peak}$.
The energy threshold can be calculated using $\eta_{th}$ through the following equation derived from \eqref{eq:2}:

\begin{equation}
  x_{th} = x_{peak} 10^{\frac{\eta_{th}}{10}}.
  \label{eq:4}
\end{equation}

For our adversarial attack on speech, $\eta_{th}$ value of -20 was suitable.
The binary mask and masked features are generated in the following manner:

\begin{equation}
  \mu[m,k] =  
  \begin{dcases}
    1, & x[m,k] \ge x_{th},   \\
    0, & x[m,k] < x_{th}.
  \end{dcases}
\label{eq:5}
\end{equation}

\begin{equation}
  x_{\text{sem}}[m,k] = \mu[m,k] x[m,k] 
  \label{eq:7}
\end{equation}

where $x[m,k]$ is the element of the feature on the energy $x[m,k]$ and $x_{\text{sem}}[m,k]$ represents the feature value masked in SEM.
Through this SEM algorithm, we can obtain the masked energy perturbation (MEP) by multiplying the previously acquired perturbations $\delta$ from adversarial attacks with the generated mask.
This process effectively applies the perturbations only to the selected time-frequency regions determined by the energy-based masking scheme.

\begin{equation}
x_{\text{mep}} = \delta \cdot x_{\text{sem}}[m,k]
\label{eq:10}
\end{equation}

where $\tilde{x}$ is generated by adding MEP to the original power spectrum $x$, and then $\tilde{x}_{n+1}$ is iteratively updated by adding MEP to $\tilde{x}_n$ as follows:

\begin{figure}[h!]
  \centering
  \includegraphics[width=1\linewidth]{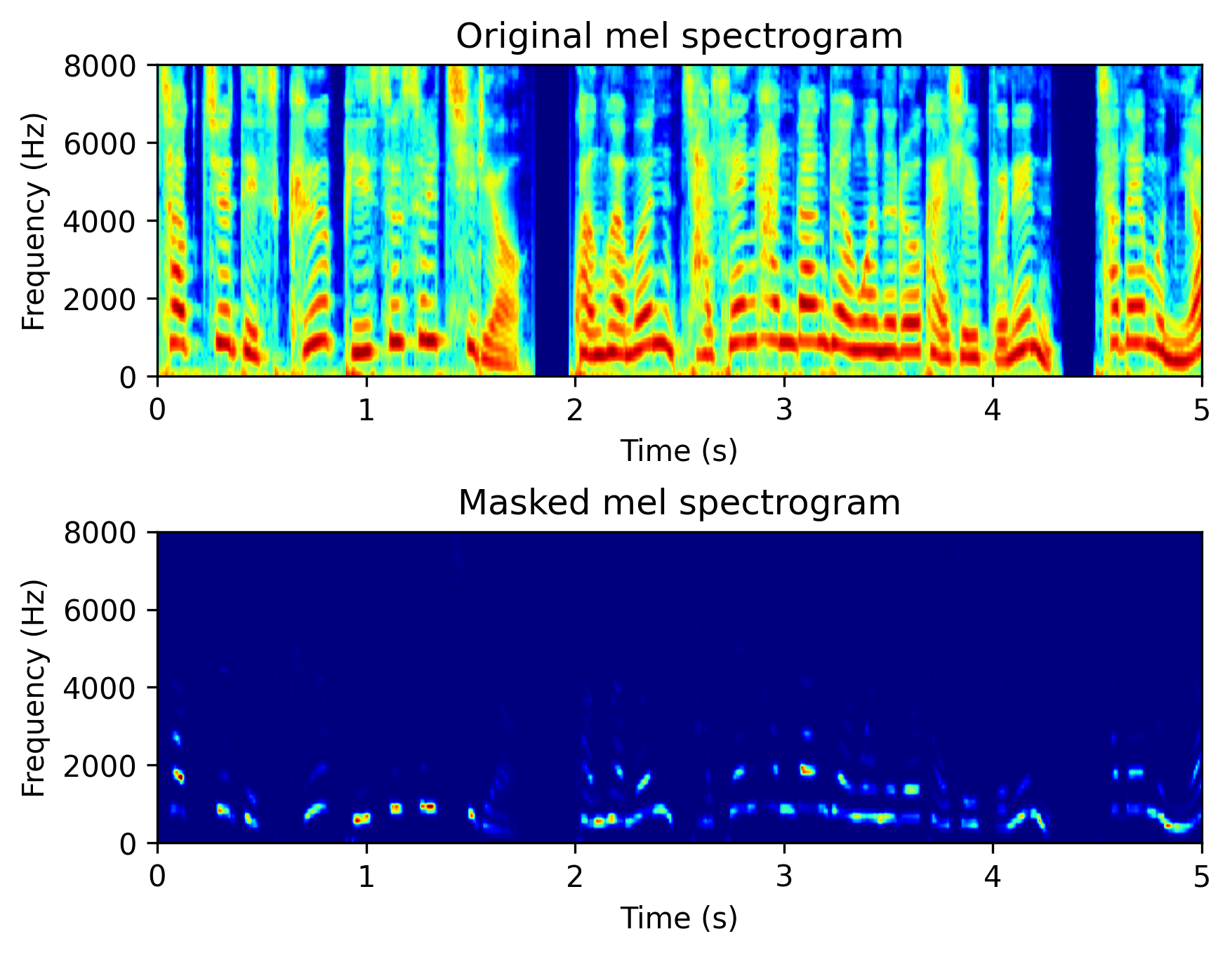}
  \caption{Mel spectrograms of a single utterance.
  The top graph shows the original speech mel spectrogram, and the bottom graph shows the mel spectrogram with the small energy region masked.}
  \label{fig:1}
\end{figure}

\begin{equation}
  \begin{aligned}
    \tilde{x}_n = x + x_{\text{mep}} \\
    \tilde{x}_n + x_{\text{mep}} \rightarrow \tilde{x}_{n+1}
  \end{aligned}
\end{equation}

The log mel spectrogram with MEP applied, which retains high energies including the 95th percentile of the original energy while masking small energies, is shown in Figure~\ref{fig:1}.

\subsection{Evasion Attack}

\begin{figure}[h!]
  \centering
  \includegraphics[width=0.85\linewidth]{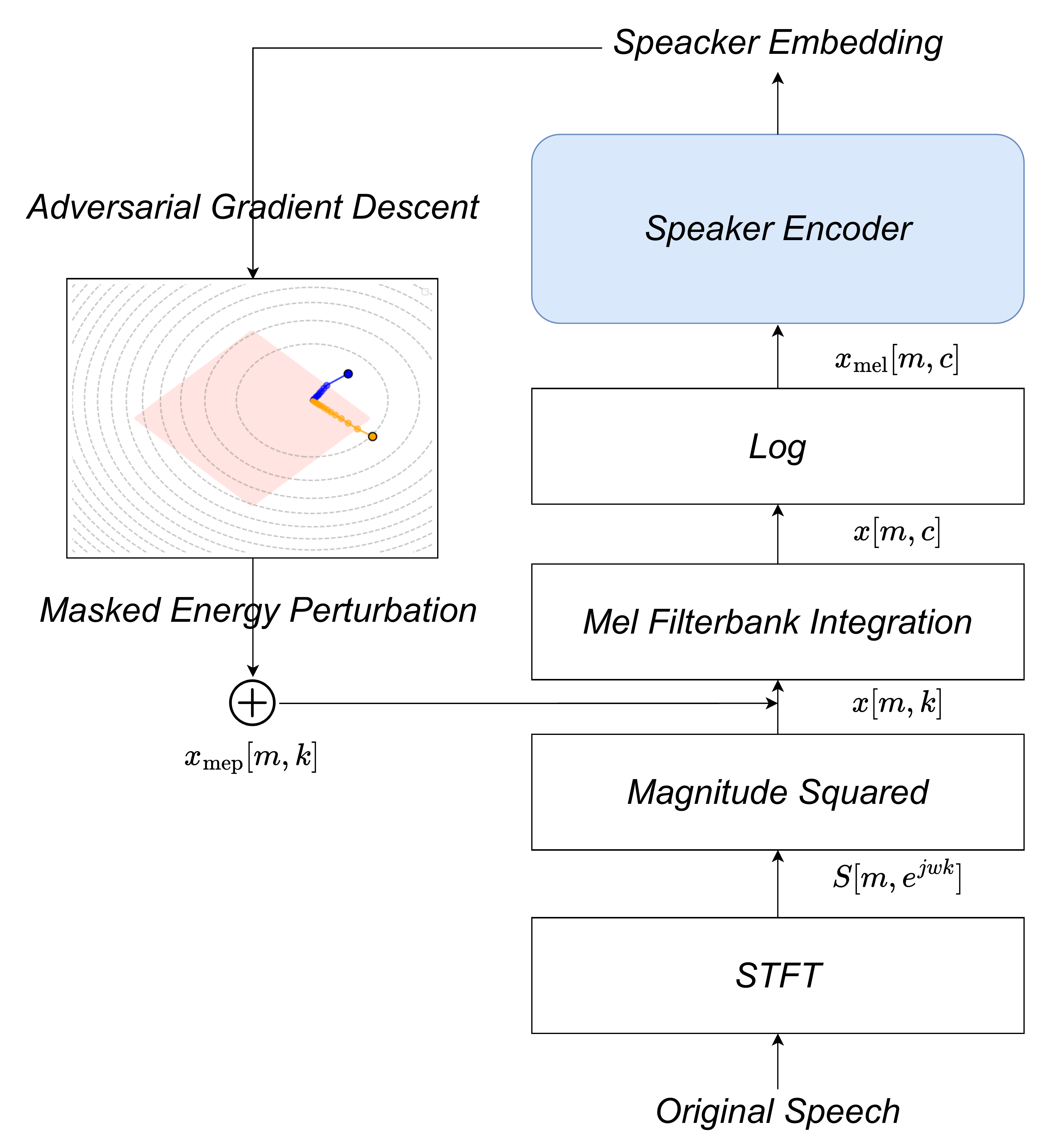}
  \caption{Block diagram for generating adversarial samples based on gradient descent with masked energy perturbation (MEP). Here, $c$ is the filterbank channel index.}
  \label{fig:2}
\end{figure}

Security in AI is a critical concern, and as the saying goes, "attack is the best defense" - evasion attacks have emerged as one of the most effective methods to understand and address these vulnerabilities.
Evasion attacks manipulate input data to an AI model to cause misclassification and induce incorrect predictions or decisions \cite{alshahrani2022adversarial}.
Occurring during the testing or deployment phase, these attacks exploit model vulnerabilities without altering the training process, making them particularly challenging to detect and mitigate.
Figure~\ref{fig:2} demonstrates the generation of adversarial samples using gradient descent based on the MEP.

Algorithm~\ref{alg:1} presents our Masked Energy Perturbation (MEP) methodology, which consists of two main attack strategies for speaker recognition systems.
The first part introduces the basic MEP attack, which processes the energy through a mel filterbank, computes embeddings using the speaker recognition system, and generates perturbations based on the gradient of the loss function masked by feature.
The second part details the Iterative MEP (I-MEP) attack, which initializes the perturbation at zero and iteratively refines it over N iterations, applying the mel filterbank transformation at each step while maintaining perturbation constraints through clipping.
Both approaches aim to create adversarial examples by adding carefully crafted perturbations to the original input features, ultimately returning an adversarial example  that combines the original input $x$ with the optimized perturbation to potentially deceive speaker recognition systems.

\begin{algorithm}[h!]
\caption{Masked Energy Perturbation}
\label{alg:1}
\begin{algorithmic}[1]
\Require energy $x$, target speaker embedding $y$, speaker recognition system $f(\cdot)$, perturbation $\delta$, loss function $L(\cdot)$, hyperparameters $\epsilon$ and $\alpha$, number of iterations $N$, masked feature $\mu$.
\State \# 1. MEP attack
\State $x_{\text{mel}} \gets \text{MelFilterBank}(x)$
\State $x_{\text{log}} \gets \log(x_{\text{mel}})$
\State $e \gets f(x_{\text{log}})$
\State $\text{loss} \gets L(e, y)$
\State $\text{masked loss} \gets L(e, y) \odot \mu$
\State $\delta \gets \alpha \cdot \mathrm{sign}(\nabla L(e, y))\odot \mu$
\State $\widetilde{x} \gets x + \delta$

\State \# 2. Iterative MEP attack
\State Initialize $\delta = 0$
\For{$i=0$ to $N-1$}
    \State $x_{\text{mel}} \gets \text{MelFilterBank}(x + \delta)$
    \State $x_{\text{log}} \gets \log(x_{\text{mel}})$
    \State $e \gets f(x_{\text{log}})$
    \State $\text{loss} \gets L(e, y)$
    \State $\text{masked loss} \gets L(e, y) \odot \mu$
    \State $\delta \gets \delta + \alpha \cdot \mathrm{sign}(\nabla L(e, y)) \odot \mu$
    \State Clip $\|\delta\| \leq \epsilon$
    \State $\widetilde{x} \gets x + \delta$
\EndFor
\State \Return adversarial waveform example $\gets \text{InverseSTFT}(\sqrt{\widetilde{x}})$
\end{algorithmic}
\end{algorithm}

\section{Experiment results and analysis}
\label{sec:setup}

\subsection{Experimental Setup}
The experiment is conducted on the LibriSpeech dataset \cite{panayotov2015librispeech} using an average of 113 utterances per speaker, randomly selected from 116 speakers in the clean set.
These recordings are derived from audiobook readings, sampled at 16 kHz, and have been carefully segmented and aligned.
The selected speech samples feature minimal background noise and are particularly suitable for speech recognition tasks due to their superior recording quality.

\subsection{Speaker Encoder Model}

We utilized pre-trained speaker recognition models and conducted experiments under white-box conditions.
The system architecture employs ResNetSE34-L, ResNetSE34-V \cite{heo2020clova}, and ECAPA-TDNN \cite{desplanques2020ecapa} models, all of which were trained on VoxCeleb \cite{nagrani2017voxceleb, chung2018voxceleb2} and operate with filterbank features as input representations.

\subsection{Attack Methods}

We compare the performance using adversarial attack methods including FGSM \cite{goodfellow2014explaining}, I-FGSM \cite{kurakin2016adversarial}, MI-FGSM \cite{dong2018boosting}, PGD \cite{madry2017towards}, MEP abd I-MEP on the speaker encoder.
All comparative methods were performed with either their optimal settings or recommended default configurations, specifically perturbation $\epsilon$ = 0.0002, iteration $N$ = 20, and step size $\alpha$ = $\epsilon/N$.

\begin{table}[htb!]
  \centering
  \caption{Performance comparison of the masked energy perturbation (MEP) method and different attack methods on a speaker encoder model.}
  \label{tab:1}
    \begin{tabular}{c|ccc}
      \hline
      \textbf{Model}                                   & \textbf{Attack Method}  & PESQ ↑ &  SNR (dB) ↑  \\
      \hline
      \multirow{6}{*}{\centering ResNet34-L}  & FGSM                    & 2.6062     &  27.58          \\
                                              & I-FGSM                  & 3.2400       &  33.49       \\
                                              & MI-FGSM                  & 3.6888       &  34.92      \\
                                              & PGD                     & 2.8424       &  29.89       \\
                                              & MEP                    & 3.6779        &   33.95      \\
                                              & \textcolor{blue}{\textbf{\textit{I-MEP}}}  & \textcolor{blue}{\textbf{3.7657}}  &  \textcolor{blue}{\textbf{38.07}}  \\
      \hline
      \multirow{6}{*}{\centering ResNet34-V}   & FGSM                    & 2.6062    &  27.60        \\
                                            & I-FGSM                  & 3.2904         &  33.94   \\
                                            & MI-FGSM                  & 3.6939       &  34.92         \\
                                            & PGD                     & 2.8425        &  29.89  \\
                                            & MEP                    &  3.6764  &  33.92    \\
                                            & \textcolor{blue}{\textbf{\textit{I-MEP}}}                     & \textcolor{blue}{\textbf{3.7709}} &  \textcolor{blue}{\textbf{38.14}}           \\
      \hline
      \multirow{6}{*}{\centering\makecell{ECAPA\\TDNN}}     & FGSM                    & 2.6115   & 27.66        \\
                                            & I-FGSM                  & 3.2815        & 33.82     \\
                                            & MI-FGSM                  & 3.6921       & 34.79        \\
                                            & PGD                     & 2.8440        & 29.87     \\
                                            & MEP                    & 3.6801         & 33.94     \\
                                            & \textcolor{blue}{\textbf{\textit{I-MEP}}}                     & \textcolor{blue}{\textbf{3.7692}}   & \textcolor{blue}{\textbf{38.09}}            \\
      \hline
    \end{tabular}
\end{table}

\subsection{Results of using MEP and attack methods on the speaker encoders}

We employed PESQ to assess the level of distortion in the distorted samples for voice quality.
Table~\ref{tab:1} shows a comparison of the effects of the Masked Energy Perturbation (MEP) technique and various attack methods (FGSM, I-FGSM, MI-FGSM, PGD, MEP, I-MEP) on three speaker encoder models (ResNetSE34-L, ResNetSE34-V, ECAPA-TDNN) using PESQ and SNR metrics.
For all three speaker encoder models, the I-MEP attack method achieves the highest PESQ scores (from 3.7657 to 3.7709) and SNR values (from 38.07 to 38.14 dB).
This indicates that I-MEP causes minimal degradation in audio quality and maintains an excellent signal-to-noise ratio.
Overall, the MEP-based methods consistently outperform traditional attacks such as FGSM, I-FGSM, MI-FGSM, and PGD.
In particular, I-MEP achieves the best performance across all three models, demonstrating that it is an effective attack method that preserves audio quality to the greatest extent when targeting speaker encoder models.
These results suggest that, compared to conventional methods like FGSM, I-FGSM, MI-FGSM, and PGD, the MEP and I-MEP attacks are better at maintaining speech signal quality.

\begin{table}[htb!]
  \centering
  \caption{Performance comparison of EER (\%) for speaker encoder models under mask energy perturbation and various attack methods.
  The baseline refers to the measurement obtained from pure inference without any adversarial attacks.}
  \label{tab:3}
    \begin{tabular}{c|cc}
      \hline
      \textbf{Model}                                   & \textbf{Attack Method}  & EER (\%) ↑   \\
      \hline
      \multirow{7}{*}{\centering ResNet34-L}  & baseline                    &   0.25             \\
                                              & FGSM                    &   41.67             \\
                                              & I-FGSM                  &   42.12              \\
                                              & MI-FGSM                  &    42.78         \\
                                              & PGD                     &  43.48               \\
                                              & MEP                    &  42.62               \\
                                              & \textcolor{blue}{\textbf{\textit{I-MEP}}}                     &  \textcolor{blue}{\textbf{44.04}}           \\
      \hline
      \multirow{7}{*}{\centering ResNet34-V}   & baseline                    & 0.01          \\
                                            & FGSM                    & 41.67          \\
                                            & I-FGSM                  &    42.29         \\
                                            & MI-FGSM                  &    40.40         \\
                                            & PGD                     &    40.06        \\
                                            & MEP                    & 42.62   \\
                                            & \textcolor{blue}{\textbf{\textit{I-MEP}}}                    &  \textcolor{blue}{\textbf{42.64}}            \\
      \hline
      \multirow{7}{*}{\centering ECAPA-TDNN}     & baseline                    & 6.38             \\
                                             & \textcolor{blue}{\textbf{\textit{FGSM}}}                    & \textcolor{blue}{\textbf{43.16}}             \\
                                            & I-FGSM                 & 42.36            \\
                                            & MI-FGSM                  &    42.29         \\
                                            & PGD                    & 41.66            \\
                                            & MEP                    & 43.37            \\
                                            & I-MEP              & 42.37                 \\
      \hline
    \end{tabular}
\end{table}

To evaluate the speaker similarity between generated adversarial speech utterances, we conducted an automatic speaker verification (ASV) assessment using the equal error rate (EER) as the performance metric.
The experimental setup included three categories: enroll, test (utterances from the same speaker as enroll but with no overlap), and imposter (utterances from different speakers).
Table~\ref{tab:3} presents a performance comparison of EER for three speaker encoder models under MEP and various adversarial attack methods.
Baseline refers to the performance evaluation of a model performed on the dataset without applying an adversarial attack.
Our MEP and I-MEP are energy masking-based attacks that explore different dimensions of vulnerability.

\section{Conclusions}
\label{sec:conclusion}

The study introduces a masked energy perturbation (MEP) approach that leverages human auditory masking phenomena to protect voice privacy.
By targeting low-energy regions in the mel-filterbank domain, MEP applies imperceptible perturbations that reduce speaker embedding similarities while preserving audio quality.
This method effectively balances privacy and perceptual naturalness by operating within psychoacoustic constraints.

\section{Acknowledgment}
his work was supported by Institute of Information \& communications Technology Planning \& Evaluation (IITP) under the artificial intelligence star fellowship support program to nurture the best talents (IITP-2025-RS-2025-02304828) grant funded by the Korea government(MSIT).

\bibliographystyle{IEEEbib}
\bibliography{strings,refs}

\end{document}